\documentclass[twocolumn,pra,aps,superscriptaddress,showpacs]{revtex4-1}
\usepackage[english]{babel}
\usepackage{amsmath}
\usepackage{graphicx}
\usepackage{amssymb}
\usepackage{txfonts}
\usepackage[unicode=true,pdfusetitle,
 bookmarks=false,bookmarksnumbered=false,bookmarksopen=false,
 breaklinks=false,pdfborder={0 0 1},backref=false,colorlinks=false]
 {hyperref}

\makeatother

\begin{document}

\title{All-optical high-resolution magnetic resonance using a nitrogen-vacancy
spin in diamond }

\author{Zhen-Yu Wang}
\affiliation{Institut f\"ur Theoretische Physik, Albert-Einstein-Allee 11, Universit\"at Ulm, 89069 Ulm, Germany}
\author{Jian-Ming Cai}
\affiliation{Institut f\"ur Theoretische Physik, Albert-Einstein-Allee 11, Universit\"at Ulm, 89069 Ulm, Germany}
\author{Alex Retzker}
\affiliation{Racah Institute of Physics, The Hebrew University of Jerusalem, Jerusalem, 91904, Israel}
\author{Martin B. Plenio}
\affiliation{Institut f\"ur Theoretische Physik, Albert-Einstein-Allee 11, Universit\"at Ulm, 89069 Ulm, Germany}

\begin{abstract}
We propose an all-optical scheme to prolong the quantum coherence
of a negatively charged nitrogen-vacancy (NV) center in diamond. Optical
control of the NV spin suppresses energy fluctuations of the $^{3}\text{A}_{2}$
ground states and forms an energy gap protected subspace. By optical control, the spectral linewidth of magnetic resonance is much narrower and the measurement of the frequencies of magnetic field sources has higher resolution.
The optical control also improves the sensitivity of the magnetic
field detection and can provide measurement of the directions of signal
sources.
\end{abstract}


\maketitle

\hyphenpenalty 1800

\section{Introduction.}

High-resolution magnetic resonance is currently one of the most important
tools in many areas of science and technology, including analytical chemistry, materials science,
structural biology, neuroscience, and medicine~\cite{Ernst:1994:OxfordUniversityPress}.
However, the sensitivity of conventional techniques is restricted to
large spin ensembles, which currently limits spatial resolution to the
micrometer scale~\cite{Ciobanu:2002:178}. Recently considerable attention
has focused on the application of negatively charged nitrogen-vacancy
(NV) centers in diamond as an atomic-sized magnetic field sensor to
detect nuclear magnetic resonance (NMR) signals by quantum control
with both laser fields and microwaves~\cite{Degen:2008:243111,Maze:2008:644,Balasubramanian:2008:648,Zhao:2011:242,Zhao:2012:657,Mamin:2013:557,Staudacher:2013:561,London:2013:067601,Muller2013Draft,Cai:2013:013020}.
In these works,
the NV centers are initialized and readout by optical fields~\cite{Doherty:2013:1,Dobrovitski:2013:23};
but the noise protection is based on pulsed~\cite{Hahn:1950:580,Carr:1954:630,Naydenov:2011:081201,Yang:2010:2,Lidar:2013:QEC} and continuous~\cite{Cai:2012:113023,cai2012long} dynamical decoupling techniques
employing microwave control. Increasing the pulse rates in the case of pulsed dynamical decoupling and their Rabi frequencies in the case of continuous dynamical decoupling beyond the GHz regime is highly challenging. The requirement
of realizing microwaves (which have wavelengths of centimeters)
imposes limitations on the setup and individual microwave control on NV centers
is difficult. Hence there has been a major effort to desirable to develop methods to overcome these shortcomings. All-optical
control was recently shown to be possible~\cite{Yale:2013:7595},
and an all-optical scheme for sensing the amplitudes of magnetic fields
was demonstrated by electromagnetically induced transparency in an
NV ensemble~\cite{Acosta:2013:213605}. However to date, there are no all-optical
methods to measure the frequencies of the magnetic fields which provides
rich magnetic resonance information about the signal sources.

In this work, we propose an all-optical magnetic resonance scheme
using a negatively charged NV center to measure the frequencies of
magnetic fields [see Fig.~\ref{fig:FigSketch}(a)]. Unlike the magnetic resonance supported by dynamical decoupling,
in our scheme fluctuations in optical control do not broaden the
resonant signal peaks, and the frequency of magnetic resonance is
determined by the energy gap of the $^{3}\text{A}_{2}$ ground sublevels,
which can easily extend the sensing frequencies to the GHz range.
The optical control of the NV center suppresses the energy fluctuations
of the $^{3}\text{A}_{2}$ ground sublevels and significantly extends
the coherence times of the NV centers. Since the magnetic resonance
linewidth broadening by dephasing is eliminated through the optical control,
high-resolution magnetic resonance with NV centers becomes possible. The
all-optical magnetic resonance scheme may also have applications in
solid-state GHz frequency standards and in all-optical quantum information
processing with NV centers.

\section{\emph{A negatively charged NV center under optical control}}

In applications of negatively charged NV centers in quantum technologies,
it is important to prolong the quantum coherence of the $^{3}\text{A}_{2}$
triplet spin ground states $|\pm1_{g}\rangle=|E_{0}\rangle|\pm1\rangle$
and $|0_{g}\rangle=|E_{0}\rangle|0\rangle$, where $|E_{0}\rangle$ refers to the orbital
state with 0 orbital angular momentum projections along the NV axis. 
The relaxation time of an NV spin can approach 200~s at low temperatures (10 K)~\cite{Jarmola:2012:197601},
whereas the dephasing time is relatively short, with typical values for the inhomogeneous 
dephasing time $T_{2}^{*}$ of 0.5 to
5~$\mu\text{s}$~\cite{Dobrovitski:2013:23,Doherty:2013:1}.
With a large energy difference between $|0_{g}\rangle$ and $|\pm1_{g}\rangle$, dephasing is the main source of decoherence and limits the overall decoherence time.

\begin{figure}
\includegraphics[width=3.4039in]{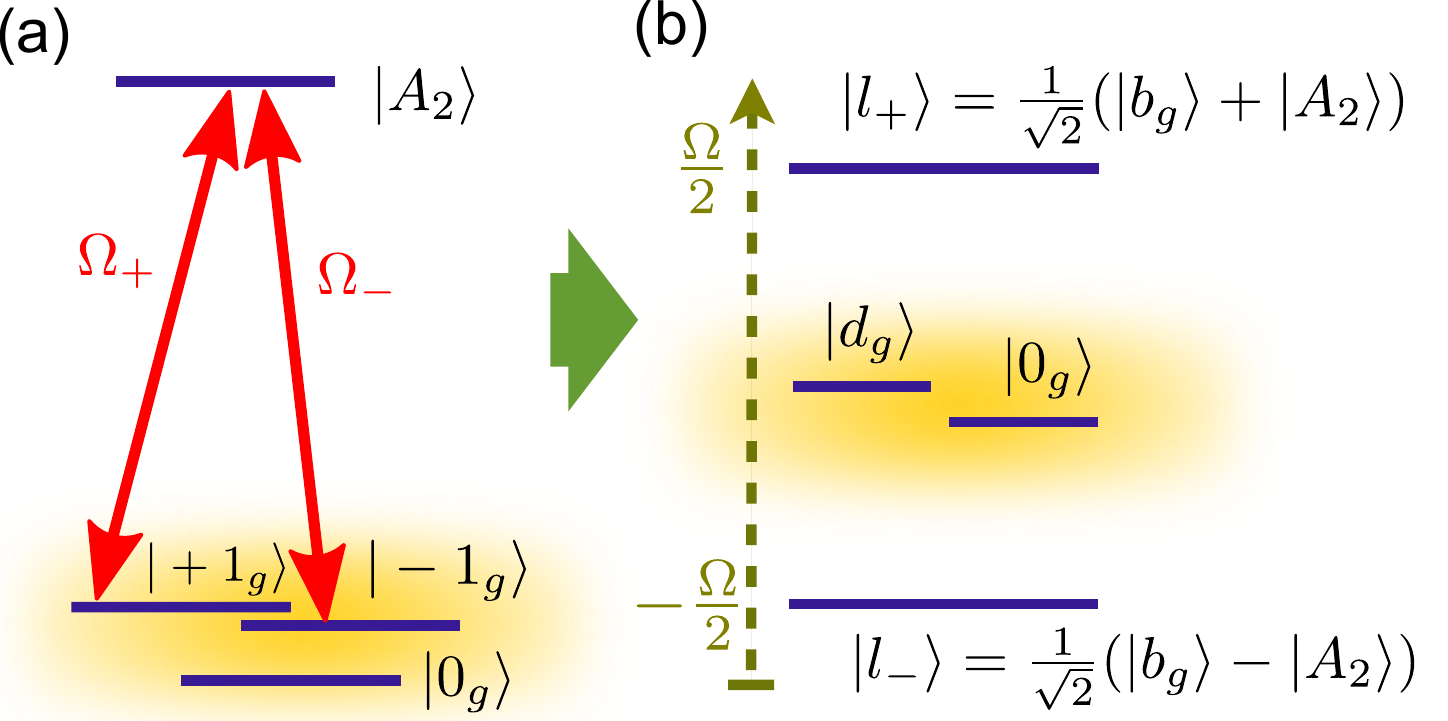}\caption{\label{fig:FigSketch}(color online). (a) NV level structure with optical control;
a $\Lambda$ system is formed by the lasers resonantly driving the transitions
between $|\pm1_{g}\rangle$ and $|A_{2}\rangle$. (b) Energy diagram
of a NV center in the dressed-state picture. The optical control induces
a coherence protected space spanned by $|d_{g}\rangle$ and $|0_{g}\rangle$
(in yellow shadow), which is protected by the energy gaps $\pm\frac{\Omega}{2}$.}
\end{figure}

For simplicity, we model the dephasing by magnetic field fluctuations
$\beta_{z}(t)$ on the NV axis (along $z$ direction), which couple to the NV spins through
the Zeeman interaction ($\hbar=1$)
\begin{equation}
H_{\text{dep}}=\beta_{z}(t)S_{z},
\end{equation}
with the spin operator $S_{z}=|+1_{g}\rangle\langle+1_{g}|-|-1_{g}\rangle\langle-1_{g}|$.
We assume that $\beta_{z}(t)$ has zero-mean $\overline{\beta_{z}(t)}=0$,
where the overline denotes ensemble averaging. The random field fluctuations
$\beta_{z}(t)$ induce broadening of the states $|\pm1_{g}\rangle$.
An initial state of the center spin $|\Psi(0)\rangle=\sum_{k=\pm1,0}a_{k}|k_{g}\rangle$
driven by $H_{\text{dep}}$ will evolve to $|\Psi(t)\rangle=a_{-1}e^{i\varphi(t)}|-1_{g}\rangle+a_{0}|0_{g}\rangle+a_{+1}e^{-i\varphi(t)}|+1_{g}\rangle$,
where the accumulated random phase $\varphi(t)=\int_{0}^{t}\beta_{z}(\tau)d\tau$.
The coherence between $|0_{g}\rangle$ and $|\pm1_{g}\rangle$ is
described by the average of the relative random phase factor $L_{0,\pm1}=\overline{e^{\pm i\varphi(t)}}$,
which vanishes when the random phase is large. For Gaussian noise,
$L_{0,\pm1}=\exp\left[-\frac{1}{2}\overline{\varphi(t)\varphi(t)}\right]$.

To suppress the dephasing using only optical control, we use two laser
fields resonantly coupling the triplet ground states $|\pm1_{g}\rangle$
to the $^{3}\text{E}$ excited state
\begin{equation}
|A_{2}\rangle=c_{+}|E_{-}\rangle|+1\rangle+c_{-}|E_{+}\rangle|-1\rangle,
\end{equation}
with $|c_{+}|^{2}+|c_{-}|^{2}=1$ (see Fig.~\ref{fig:FigSketch}).
The lasers also couple the states $|\pm1_{g}\rangle$ and $|A_{1}\rangle=c_{-}^{*}|E_{-}\rangle|+1\rangle-c_{+}^{*}|E_{+}\rangle|-1\rangle$
but with a large detuning $\delta$, which is the energy gap between the
states $|A_{2}\rangle$ and $|A_{1}\rangle$. The optical transitions
between $^{3}\text{A}_{2}$ and $^{3}\text{E}$ are spin conserving~\cite{Batalov:2009:195506,Togan:2010:730,Maze:2011:025025}.
The state properties of the NV centers, such as the parameters $c_{+}$ and $c_{-}$, depend on electric, magnetic, and strain fields. The effective Hamiltonian
to obtain the eigenstates and eigenenergies of the $^{3}\text{E}$
levels at low temperatures can be found in the review paper~\cite{Doherty:2013:1}. To have well-resolved excited states, we put the NV center at cryogenic
temperatures ($\lesssim10$ K).
Using $|E_{-}\rangle|+1\rangle=c_{+}^{*}|A_{2}\rangle+c_{-}|A_{1}\rangle$
and $|E_{+}\rangle|-1\rangle=c_{-}^{*}|A_{2}\rangle-c_{+}|A_{1}\rangle$,
we have the Hamiltonian under constant optical control
\begin{eqnarray}
H_{0} & = & \left(\Omega_{+}e^{i\phi_{+}}e^{i\omega_{+}t}\left[c_{+}^{*}|A_{2}\rangle+c_{-}|A_{1}\rangle\right]\langle+1_{g}|+\text{H. c.}\right)\nonumber \\
 &  & +\left(\Omega_{-}e^{i\phi_{-}}e^{i\omega_{-}t}\left[c_{-}^{*}|A_{2}\rangle-c_{+}|A_{1}\rangle\right]\langle-1_{g}|+\text{H. c.}\right)\nonumber \\
 &  & -\delta|A_{1}\rangle\langle A_{1}|+\sum_{k_{g}=\pm1_{g},0_{g}}E_{k_{g}}|k_{g}\rangle\langle k_{g}|+H_{\text{dep}}+H_{\text{sig}},
\end{eqnarray}
where $\Omega_{\pm}$ are the Rabi frequencies and $E_{k_{g}}$ are the
energies of the ground states. $\phi_{\pm}$ and $\omega_{\pm}$ are the phases and frequencies of the lasers, respectively. We also include an interaction
Hamiltonian $H_{\text{sig}}$ for possible signal sources. The energy
of $|A_{2}\rangle$ is set as the reference energy. The laser fields
resonantly drive the transitions between $|\pm1_{g}\rangle$ and $|A_{2}\rangle$
with the laser detuning $\Delta_{\pm1_{g}}=E_{\pm1_{g}}-\omega_{\pm}=0$.
In the rotating frame of $e^{-iH_{g}t}$ with
\begin{equation}
H_{g}=\sum_{k_{g}=\pm1_{g}}(E_{k_{g}}-\Delta_{k_{g}})|k_{g}\rangle\langle k_{g}|+E_{0_{g}}|0_{g}\rangle\langle0_{g}|,
\end{equation}
the system Hamiltonian reads
\begin{equation}
H=H_{L}+H_{\text{dep}}+\tilde{H}_{\text{sig}},\label{eq:H}
\end{equation}
where
\begin{eqnarray}
H_{L} & = & \left(\Omega_{+}e^{i\phi_{+}}\left[c_{+}^{*}|A_{2}\rangle+c_{-}|A_{1}\rangle\right]\langle+1_{g}|+\text{H. c.}\right)-\delta|A_{1}\rangle\langle A_{1}|\nonumber \\
 &  & +\left(\Omega_{-}e^{i\phi_{-}}\left[c_{-}^{*}|A_{2}\rangle-c_{+}|A_{1}\rangle\right]\langle-1_{g}|+\text{H. c.}\right),\label{eq:H_L}
\end{eqnarray}
\begin{equation}
\tilde{H}_{\text{sig}}=e^{iH_{g}t}H_{\text{sig}}e^{-iH_{g}t}.
\end{equation}

For accurate numerical simulations, we model the NV spin with
6 levels: three ground states $|\pm1_{g}\rangle$ and $|0_{g}\rangle$,
the two excited states $|A_{1}\rangle$ and $|A_{2}\rangle$, and
a singlet state $|s\rangle$ to describe the intersystem crossing
transitions. The dynamics of the NV center spin described by a density
matrix $\rho(t)$ is governed by the Lindblad master equation~\cite{Rivas:2012:OpenQuantumSystems},
\begin{equation}
\frac{d}{dt}\rho=-i\left[H,\rho\right]+\sum_{\alpha,\beta}\gamma_{\beta\alpha}\left(\sigma_{\beta\alpha}\rho\sigma_{\alpha\beta}-\frac{1}{2}\rho\sigma_{\alpha\alpha}-\frac{1}{2}\sigma_{\alpha\alpha}\rho\right),\label{eq:materEq}
\end{equation}
where the Lindblad operators $\sigma_{\beta\alpha}\equiv|\beta\rangle\langle\alpha|$, $\gamma_{\beta\alpha}$ are the decay rates, 
and the Hamiltonian $H$ is given by Eq.~(\ref{eq:H}).

\section{\emph{Noise suppression by optical control}}

To illustrate the fundamental idea of noise suppression by optical control, we
consider a simplified model without contributions from spontaneous decay (taken into account in detailed numerical simulations to present subsequently). When the energy gap $\delta\gg\Omega_{\pm}$,
the coupling to $|A_{1}\rangle$ is negligible in Eq.~(\ref{eq:H}),
and we have the $\Lambda$-type Hamiltonian by dropping out terms related
to the state $|A_{1}\rangle$,

\begin{equation}
H_{\Lambda}=H_{L2}+H_{\text{dep}}+\tilde{H}_{\text{sig}},\label{eq:HLambda}
\end{equation}
where
\begin{equation}
H_{L2}=\Omega|A_{2}\rangle\langle b_{g}|+\text{H. c.},
\end{equation}
with the effective Rabi frequency
\begin{equation}
\Omega\equiv\sqrt{\Omega_{+}^{2}|c_{+}|^{2}+\Omega_{-}^{2}|c_{-}|^{2}},
\end{equation}
and the bright state

\begin{equation}
|b_{g}\rangle\equiv\frac{1}{\Omega}\left(c_{+}\Omega_{+}e^{-i\phi_{+}}|+1_{g}\rangle+c_{-}\Omega_{-}e^{-i\phi_{-}}|-1_{g}\rangle\right).
\end{equation}
The laser driving fields form a $\Lambda$ system with an excited state
$|A_{2}\rangle$ and two ground states $|\pm1_{g}\rangle$ (see Fig.~\ref{fig:FigSketch}).
The dark state decoupled from the laser is
\begin{equation}
|d_{g}\rangle\equiv e^{i\varphi_{d}}\frac{1}{\Omega}\left(c_{-}^{*}\Omega_{-}e^{-i\phi_{+}}|+1_{g}\rangle-c_{+}^{*}\Omega_{+}e^{-i\phi_{-}}|-1_{g}\rangle\right),\label{eq:stateDg}
\end{equation}
with a global phase $\varphi_{d}=\text{Arc}(c_{+}c_{-})$. The laser Hamiltonian $H_{L2}$
has the eigenstates $|d_{g}\rangle$, $|0_{g}\rangle$, and $|l_{\pm}\rangle=\frac{1}{\sqrt{2}}\left(|b_{g}\rangle\pm|A_{2}\rangle\right)$
with energies $E_{l\pm}=\pm\frac{\Omega}{2}$. The subspace spanned
by the two states $|d_{g}\rangle$ and $|0_{g}\rangle$ are separated
from the other eigenstates $|l_{\pm}\rangle$ by the energy gaps $\pm\frac{\Omega}{2}$.
Therefore the transitions from this subspace to $|l_{\pm}\rangle$
are suppressed by an energy penalty that is proportional to the Rabi frequency of the optical driving fields  (see Fig.~\ref{fig:FigSketch}
and Ref.~\cite{Timoney:2011:185}).

The spin operator $S_{z}$ in the basis of $|b_{g}\rangle$ and $|d_{g}\rangle$ reads
\begin{eqnarray}
S_{z} & = & \frac{1}{\Omega}\left(2e^{-i\varphi_{d}}c_{+}\Omega_{+}c_{-}\Omega_{-}\right)|d_{g}\rangle\langle b_{g}|+\text{H.c.} \nonumber \\ 
 &  & +\kappa\left(|b_{g}\rangle\langle b_{g}|-|d_{g}\rangle\langle d_{g}|\right).\label{eq:SpinOperator}
\end{eqnarray}
where $\kappa=\frac{1}{\Omega}\left(|c_{+}\Omega_{+}|^2-|c_{-}\Omega_{-}|^2\right)\leq 1$.
To suppress the dephasing of the NV spin using the energy penalty $\frac{\Omega}{2}$,
we choose the laser fields that satisfy
\begin{equation}
|c_{+}\Omega_{+}|=|c_{-}\Omega_{-}|.\label{eq:OmegaRelativeAmp}
\end{equation}
Under such a condition,
the spin operator $S_{z}$ becomes
\begin{equation}
S_{z}=|b_{g}\rangle\langle d_{g}|+\text{H.c.}=\frac{1}{\sqrt{2}}\left(|l_{+}\rangle+|l_{-}\rangle\right)\langle d_{g}|+\text{H.c.},
\end{equation}
with
\begin{equation}
|b_{g}\rangle=e^{i\varphi_{b}}\frac{1}{\sqrt{2}}\left[|+1_{g}\rangle+e^{i\phi_{L}}|-1_{g}\rangle\right],\label{eq:stateBg-eq}
\end{equation}
\begin{equation}
|d_{g}\rangle=e^{i\varphi_{b}}\frac{1}{\sqrt{2}}\left[|+1_{g}\rangle-e^{i\phi_{L}}|-1_{g}\rangle\right].\label{eq:stateDg-eq}
\end{equation}
where $\varphi_{b}=\text{Arc}(c_{+}\Omega_{+}e^{-i\phi_{+}})$ and
$\phi_{L}=\text{Arc}(\Omega_{+}\Omega_{-}c_{+}^{*}c_{-})+\phi_{+}-\phi_{-}$
is a tunable relative phase. 
With a relative large $\Omega$, the spectral power density of the
magnetic field fluctuations $\beta_{z}(t)$ at the frequencies around
the energy gap $\frac{\Omega}{2}$ is negligible. The off-resonant
fluctuations $\beta_{z}(t)$ cannot induce the transitions from $|d_{g}\rangle$
to $|l_{\pm}\rangle$, which are strongly suppressed by the
energy penalty $\pm\frac{\Omega}{2}$ [see Fig.~\ref{fig:FigSketch}(b)]. Because fluctuations in the effective Rabi frequency $\Omega$ only cause small changes in the magnitudes of the energy gap, our scheme is stable against the fluctuations of $\Omega$~\cite{Timoney:2011:185}, as long as the magnitudes of the energy gap are still much larger than the fluctuation frequencies of $\beta_{z}(t)$. When there are relative fluctuations in $\Omega_{\pm}$ and $c_{\pm}$, the second line in Eq.~(\ref{eq:SpinOperator}) does not vanish, and a fraction $\sim\kappa$ of the noise will not be suppressed by the energy gaps $\pm\frac{\Omega}{2}$ made by the optical control. The decoherence time caused by this fraction of noise is estimated to be $T_{\text{frac}}\sim T_{2}^{*}/\kappa$, and the fluctuations only cause negligible effects if $T_{\text{frac}}$ is much larger than the controlled evolution time.
The relative amplitude fluctuations in the driving fields can be made very small if the fields are obtained from the same laser.

To manifest the effects of dephasing, we set $\tilde{H}_{\text{sig}}=0$
in Eq.~(\ref{eq:HLambda}). The quantum coherence between, e.g., $|0_{g}\rangle$
and $|d_{g}\rangle$ is described by the average
\begin{equation}
L_{0,d_{g}}(t)=\overline{\langle d_{g}|U_{\Lambda}(t)|d_{g}\rangle\langle0_{g}|U_{\Lambda}^{\dagger}(t)|0_{g}\rangle},
\end{equation}
where $U_{\Lambda}(t)=\mathcal{T}e^{-i\int_{0}^{t}H_{\Lambda}d\tau}$ with the time-ordering operator $\mathcal{T}$.
The coherence $L_{0,d_{g}}(t)$ can be simplified as
\begin{eqnarray}
L_{0,d_{g}}(t) & = & \overline{\langle d_{g}|\tilde{U}_{\text{dep}}(t)|d_{g}\rangle},
\end{eqnarray}
by using the transformation $\tilde{U}_{\text{dep}}(t)\equiv e^{iH_{L2}t}U_{\Lambda}(t)=\mathcal{T}e^{-i\int_{0}^{t}\tilde{H}_{\text{dep}}(\tau)d\tau}$
with
\begin{flalign}
\tilde{H}_{\text{dep}}(t) & =e^{iH_{L2}t}H_{\text{dep}}e^{-iH_{L2}t}\\
& = \beta_{z}(t)\left(\cos\frac{\Omega t}{2}|b_{g}\rangle-i\sin\frac{\Omega t}{2}|A_{2}\rangle\right)\langle d_{g}|+\text{H.c.}.\label{eq:TildeHdep}
\end{flalign}
The coherence between $|0_{g}\rangle$ and $|d_{g}\rangle$ decreases
when the absolute value of $L_{0,d_{g}}(t)$ decreases. Without optical
driving fields, $L_{0,d_{g}}(t)=\exp\left[-\frac{1}{2}\overline{\left(\int_{0}^{t}\beta_{z}d\tau\right)^{2}}\right]$
for Gaussian noise and vanishes when the random phase is large. From
the expression of Eq.~(\ref{eq:TildeHdep}), we can see that when
the noise is relatively slow compared to the frequency $\frac{\Omega}{2}$,
the effect of the noise $\beta_{z}(t)$ is averaged out by the oscillating
functions $\cos\frac{\Omega t}{2}$ and $\sin\frac{\Omega t}{2}$.
Dynamical decoupling also uses similar modulation functions to average out
unwanted noise~\cite{Yang:2010:2,Lidar:2013:QEC,Cai:2012:113023,cai2012long}. A more detailed analysis in frequency domain is given in Appendix~\ref{app:freqDomain}.

\begin{figure}
\includegraphics[width=3.4039in]{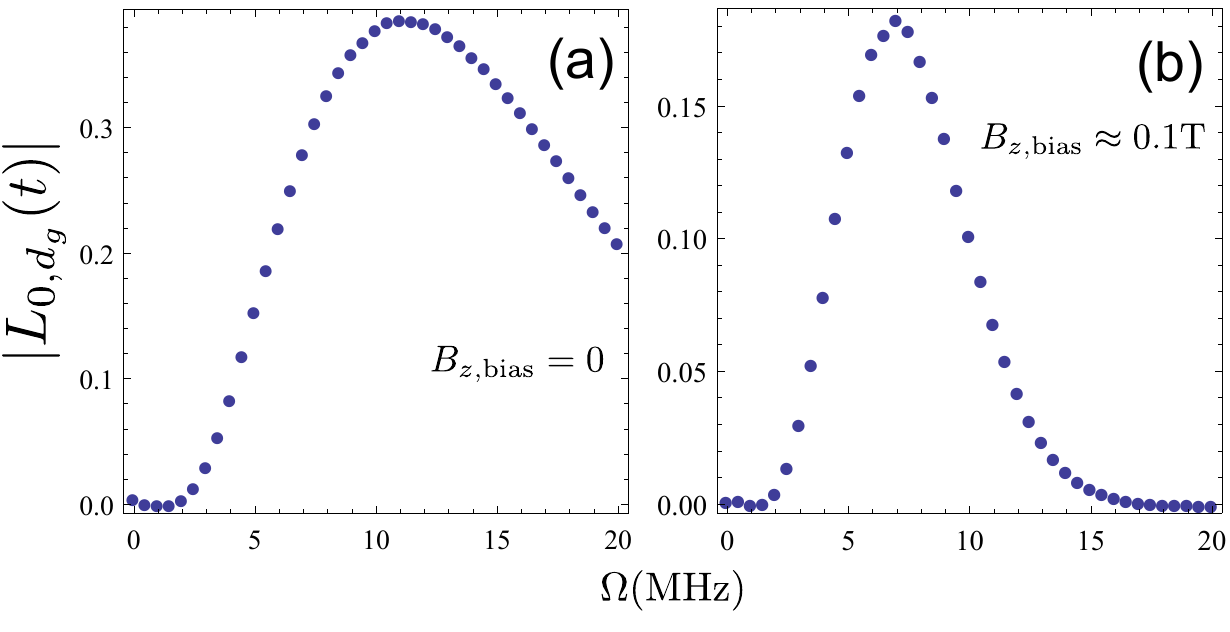}\caption{\label{fig:FigCoherVsCtr}(color online). Quantum coherence between states $|d_{g}\rangle$
and $|0_{g}\rangle$ at the moment $t=50~\mu$s as a function $\Omega$,  for  (a) $B_{z,\text{bias}}=0$ and  (b ) $B_{z,\text{bias}}\approx0.1$~T. The data were obtained by $4\times10^4$ runs of averaging. }
\end{figure}

To demonstrate our scheme, we performed numerical simulations by implementing the master equation~(\ref{eq:materEq}), including the spontaneous decay and dephasing.
We used the
parameters in the experimental paper~\cite{Yale:2013:7595} for the numerical simulation.
The decay rate from the excited states $|A_{1}\rangle$ and $|A_{2}\rangle$
to the ground sates $|\pm1_{g}\rangle$ and $|0_{g}\rangle$ is $\gamma_{g,e}\approx17$~MHz;
the rate for intersystem crossing from the excited states to the singlet
$|s\rangle$ is $\gamma_{s,e}\approx37$~MHz; the inverse intersystem
crossing rate from $|s\rangle$ to the ground states is $\gamma_{g,s}\approx2.7$~MHz.
In the simulation, the noise fluctuations $\beta_{z}(t)$ were simulated
by the Ornstein-Uhlenbeck process~\cite{Wang:1945:323},
which is Gaussian. Generated by the Ornstein-Uhlenbeck
processes, the expectation value of $\beta_{z}(t)$ is $\overline{\beta_{z}(t)}_{\text{OU}}=\beta_{z}(t_{0})e^{-(t-t_{0})/\tau_{\beta}}$;
the two-point correlation $\overline{\beta_{z}(t)\beta_{z}(t^{\prime})}_{\text{OU}}=\frac{c_{\beta}\tau_{\beta}}{2}e^{-|t-t^{\prime}|\tau_{\beta}}(1-e^{-2(t-t_{0})/\tau_{\beta}})$,
where $t_{0}$ is the starting time to generate an Ornstein-Uhlenbeck
process and $c_{\beta}$ is a diffusion coefficient~\cite{Bibbona:2008:S117}.
These quantities converge to stationary values after a time larger
than the correlation time $\tau_{\beta}$. In the simulation, we chose
$\beta_{z}(t_{0})=0$ for $\overline{\beta_{z}(t)}=0$. We used $t_{0}=-10\tau_{\beta}$ and simulated the spin dynamic at
$t\geq0$, so that $e^{-2(t-t_{0})/\tau_{\beta}}\approx0$ and we had an exponentially-decaying correlation function $\overline{\beta_{z}(t)\beta_{z}(t^{\prime})}\approx\frac{c_{\beta}\tau_{\beta}}{2}e^{-|t-t^{\prime}|\tau_{\beta}}$. At $t\ge0$, $\beta_{z}(t)$ can be treated as stationary
stochastic process. We chose the correlation time $\tau_{\beta}=25$~$\mu\text{s}$ in the simulation~\cite{de2010universal}.
The realizations of Ornstein-Uhlenbeck processes were generated
by the exact simulation algorithm in Ref.~\cite{Gillespie:1996:2084},
\begin{equation}
\beta(t+\Delta t)=\beta(t)e^{-\Delta t/\tau_{\beta}}+n_{\text{G}}\sqrt{\frac{c_{\beta}\tau_{\beta}}{2}(1-e^{-2\Delta t/\tau_{\beta}})},\label{eq:OUupdate}
\end{equation}
which requires the generation of a unit Gaussian random number $n_{\text{G}}$
at each time step. The algorithm is exact because the update algorithm Eq.~(\ref{eq:OUupdate})
is valid for any finite time step $\Delta t$~\cite{Gillespie:1996:2084}.
We chose the value of the diffusion coefficient $c_{\beta}\approx4/(T_{2}^{*2}\tau_{\beta})$
so that without optical driving fields $L_{0,\pm1}(T_{2}^{*})=e^{-1}$ at
the dephasing time $T_{2}^{*}=3$~$\mu\text{s}$.

To demonstrate the coherence protection by optical control, we prepared
the quantum state in the superposition $|\Psi(0)\rangle=\frac{1}{\sqrt{2}}\left(|d_{g}\rangle+|0_{g}\rangle\right)$
of $|d_{g}\rangle$ and $|0_{g}\rangle$.
To obtain the optimal amplitude of driving fields, we plotted the coherence $|L_{0,d_{g}}(t)|$ at $t=50$ $\mu$s
at different $\Omega$ in Fig.~\ref{fig:FigCoherVsCtr}. By increasing the driving amplitude $\Omega$, the quantum coherence was recovered by suppressing the noise effects. However, increasing $\Omega$ also increases the coupling to the state $|A_1\rangle$ [see Eq.~(\ref{eq:H_L})], which degrades the approximated $\Lambda$-type system [see Eq.~(\ref{eq:HLambda})] and leads to population leakage out of the subspace spanned by $|d_{g}\rangle$ and $|0_{g}\rangle$.
At zero static field, the energy gap $\delta\approx2$ GHz and the mixing coefficients $|c_{+}|=|c_{-}|$ in Eq.~(\ref{eq:H_L}), and therefore $|d_{g}\rangle$ only directly couples to $|A_1\rangle$.
For $\Omega\ll\delta$, the population to $|A_1\rangle$ state is $\sim\left({\Omega}/{\delta}\right)^2$, estimated by time-independent perturbation theory. The value of $L_{0,d_{g}}(t)$ at a moment $t$ decreases when the leakage $\sim\left({\Omega}/{\delta}\right)^2\gamma_{e}t$ increases. Here $\gamma_{e}$ is the estimated decay rate from the excited states. In Fig.~\ref{fig:FigCoherVsCtr}(a), the optimal control with $\Omega\sim10$ MHz gives the best coherence protection. We can increase $\delta$ by applying a static bias magnetic field $B_{z,\text{bias}}$ along the axis of the NV center (z~direction). However, non-zero $B_{z,\text{bias}}$ also changes the mixing coefficients $c_{\pm}$. For example, $B_{z,\text{bias}}=0.1$~T gives $\delta\approx 5.71$ GHz, $c_{+}\approx0.984$, and
$c_{-}\approx0.178$. When $|c_{+}|\neq|c_{-}|$, the dark state $|d\rangle$ can transit to both excited states $|A_1\rangle$ and $|A_2\rangle$, and $|L_{0,d_{g}}(t)|$ is strongly reduced [see Fig.~\ref{fig:FigCoherVsCtr}(b)]. At $B_{z,\text{bias}}=0.1$~T, the effective Rabi frequency $\Omega\sim7$ MHz yields the best coherence protection.

\begin{figure}
\includegraphics[width=3.4039in]{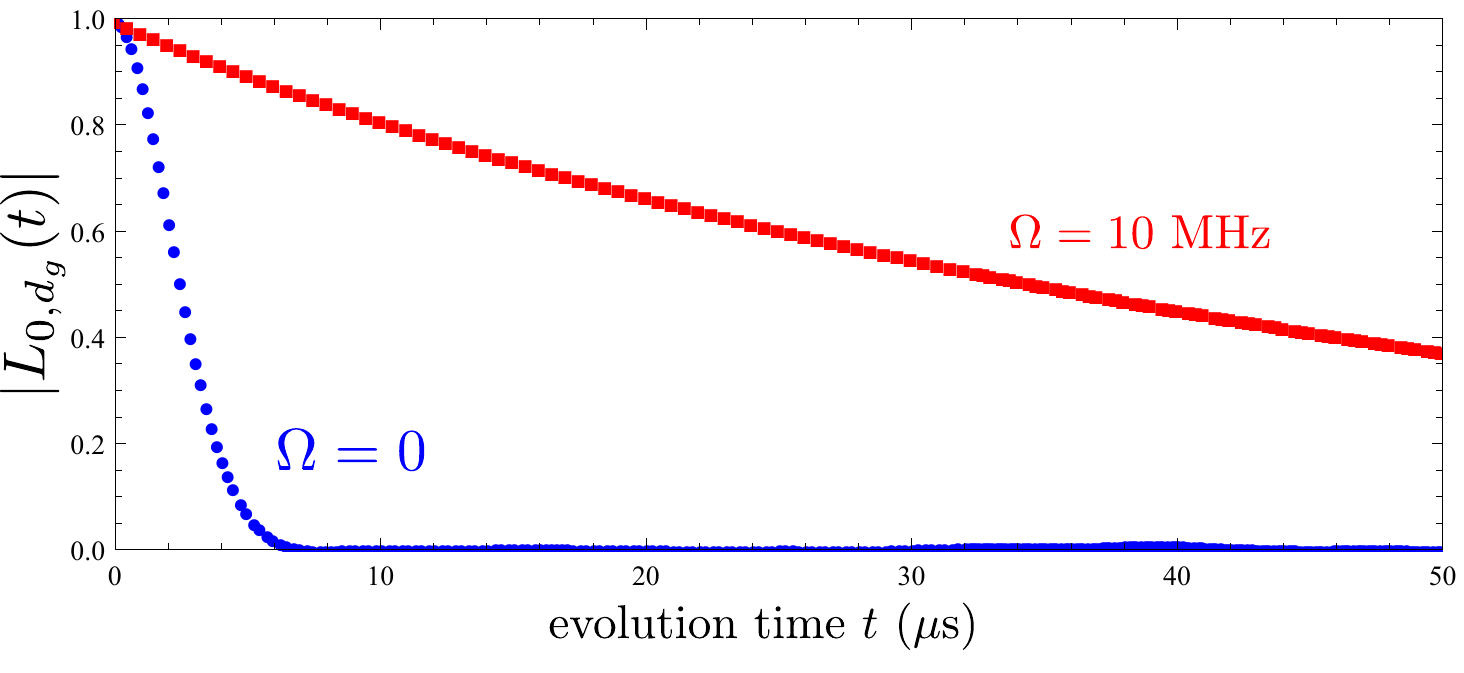}\caption{\label{fig:FigCoher}(color online). Coherence between states $|d_{g}\rangle$
and $|0_{g}\rangle$ for cases without optical control ($\Omega=0$,
blue dots) and with optical control ($\Omega=10$~MHz, red squares) at $B_{z,\text{bias}}=0$. The data were obtained by $4\times10^4$ runs of averaging.}
\end{figure}

In Fig.~\ref{fig:FigCoher}, we plot the quantum coherence as a function of time
for cases of free induction decay ($\Omega=0$) and with optical driving fields
($\Omega=10$~MHz) at zero static fields.
With an optical Rabi frequency $\Omega=10$~MHz, the coherence is
significantly prolonged and exceeds 50~$\mu$s, which is over 16-fold improvement in the coherence time $T_{2}^{*}\approx3$~$\mu$s.

\begin{figure}
\includegraphics[width=3.4039in]{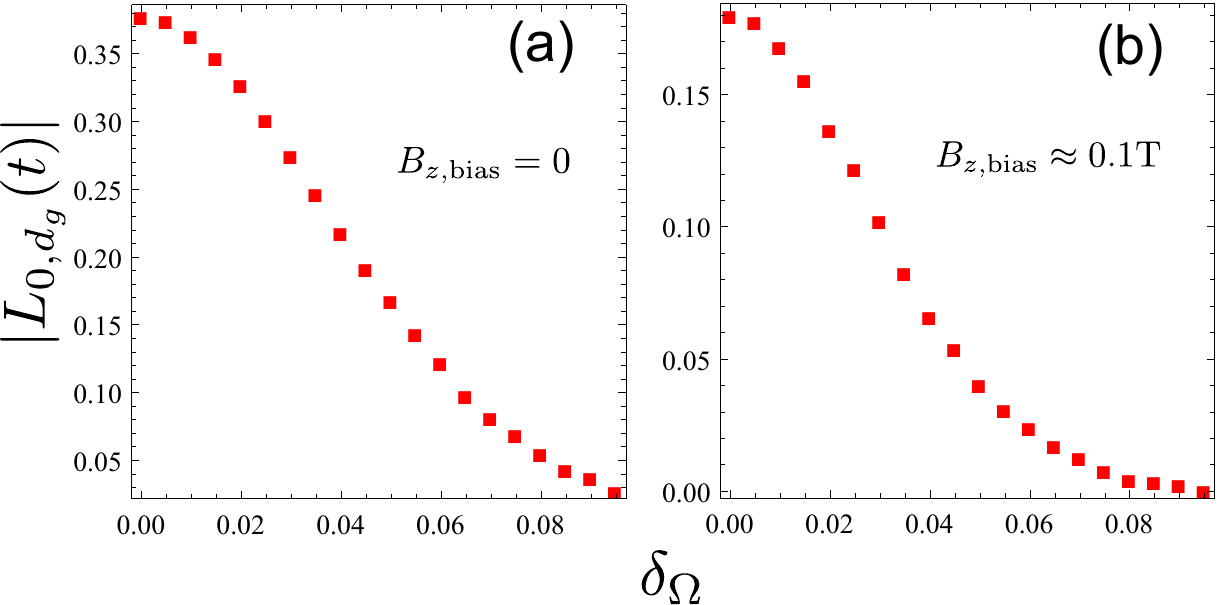}\caption{\label{fig:FigCoherVsFluct}(color online). Coherence between the states $|d_{g}\rangle$
and $|0_{g}\rangle$ as a function of the standard
deviation of relative fluctuation $\delta_{\Omega}$, at the moment $t=50~\mu$s. (a) $B_{z,\text{bias}}=0$ and $\Omega=10$~MHz;  (b) $B_{z,\text{bias}}\approx0.1$~T and $\Omega=7$~MHz. The data were obtained by $4\times10^{4}$ runs of averaging. }
\end{figure}

Although the scheme is robust to the fluctuations of $\Omega$~\cite{Timoney:2011:185}, Eq.~(\ref{eq:OmegaRelativeAmp}) shows that independent fluctuations $\delta\Omega_{\pm}(t)$ in the amplitudes of $\Omega_{\pm}$ change the bright and dark states in Eqs.~(\ref{eq:stateBg-eq}) and (\ref{eq:stateDg-eq}).
To demonstrate that the scheme is not sensitive to  independent
fluctuations $\delta\Omega_{\pm}(t)$, we modelled
$\delta\Omega_{\pm}(t)$ by Ornstein-Uhlenbeck processes. We selected a diffusion coefficient $c_{\Omega}=2\delta_{\Omega}^{2}/\tau_{\Omega}$
with a correlation time $\tau_{\Omega}=100$~$\mu$s and a variance of relative fluctuations $\overline{[\delta\Omega_{\pm}(t)/\Omega_{\pm}]^{2}}=\delta_{\Omega}^{2}$.
The impact of independent intensity fluctuations on the coherence is shown in Fig.~\ref{fig:FigCoherVsFluct}.
The scheme still provides good performance even though the standard
deviation of relative fluctuations $\delta_{\Omega}$ reaches $\sim0.02$.

\section{\emph{High-resolution magnetic resonance by optical control}}

In magnetic resonance spectroscopy, when the energy gap, e.g., $\epsilon_{0,-1}\equiv E_{0}-E_{-1}$
between $|0_{g}\rangle$ and $|-1_{g}\rangle$, coincides with the frequency
of the magnetic field, resonance transitions induce a signal peak in
the frequency domain. The linewidth and depth of the peak determine the resolution and sensitivity
of the spectroscopy. Because of the energy broadening induced by the
random fluctuations $\beta_{z}(t)$, the minimum linewidth of the
signal peak is limited by the deviation of the fluctuations $\sim\left(\overline{\beta_{z}(t)^{2}}\right)^{1/2}$.
When the dephasing (i.e., the effect of energy broadening) is suppressed,
we achieve a narrower linewidth, and hence spectroscopy with higher
accuracy.

We assume that the signal fields have negligible frequency components
around the energy gap $\frac{\Omega}{2}$. The signal Hamiltonian in Eq.~(\ref{eq:HLambda})
is written as
\begin{equation}
\tilde{H}_{\text{sig}}=\eta_{\text{sig}}(t)e^{iH_{g}t}\left[S_{x}\cos\theta_{\text{sig}}+S_{y}\sin\theta_{\text{sig}}\right]e^{-iH_{g}t},\label{eq:SignalHamiltonian}
\end{equation}
where $\eta_{\text{sig}}(t)$ is the magnetic signal field with zero
mean $\overline{\eta_{\text{sig}}(t)}=0$ and $\theta_{\text{sig}}$
is the direction of the magnetic field in the $x-y$ plane normal to the NV axis. We initialize
the NV center in the state $|0_{g}\rangle$ by optical pumping. The
population that remains in the initial state $|0_{g}\rangle$ is approximately governed by the dynamics induced by $H_{\Lambda}$ given by Eq.~(\ref{eq:HLambda}),
\begin{equation}
P_{|0_{g}\rangle}(t)=\overline{|\langle0_{g}|e^{-i\int_{0}^{t}H_{\Lambda}d\tau}|0_{g}\rangle|^{2}}.
\end{equation}

In the simulation, we generated single-frequency sources $\eta_{\text{sig}}(t)=\eta_{0}\cos(\omega_{s}t+\varphi_{s})$
with initial random phases $\varphi_{s}$ at each run of the simulation.
To have the accurate resonance frequency $\omega_{\text{Res}}$, we
diagonalized the Hamiltonian $H$ given by Eq.~(\ref{eq:H_L}), as
$\omega_{\text{Res}}$ is the energy between $|0_{g}\rangle$ and
the state $|\tilde{d}_{g}\rangle$ which is approximately $|d_{g}\rangle$
and has a little mixing with the excited states $|A_{2}\rangle$ and
$|A_{1}\rangle$. For $\Omega\ll\delta$, $\omega_{\text{Res}}\approx\epsilon_{0,-1}$, up to a correction $\sim\Omega^{2}/\delta$.
We consider the magnetic resonance signal where the difference between the signal frequency and the resonant frequency is small, i.e., $|\omega_{s}-\omega_{\text{Res}}|\ll\omega_{\text{Res}}$. This enables the application of a rotating wave approximation by neglecting oscillating terms with frequencies $\sim2\omega_{\text{Res}}$ in Eq.~(\ref{eq:SignalHamiltonian}). The simulations used the master equation~(\ref{eq:materEq}) with the Hamiltonian Eq.~(\ref{eq:H}).

\subsection{Measurement of signal frequencies}

\begin{figure}
\includegraphics[width=3.4039in]{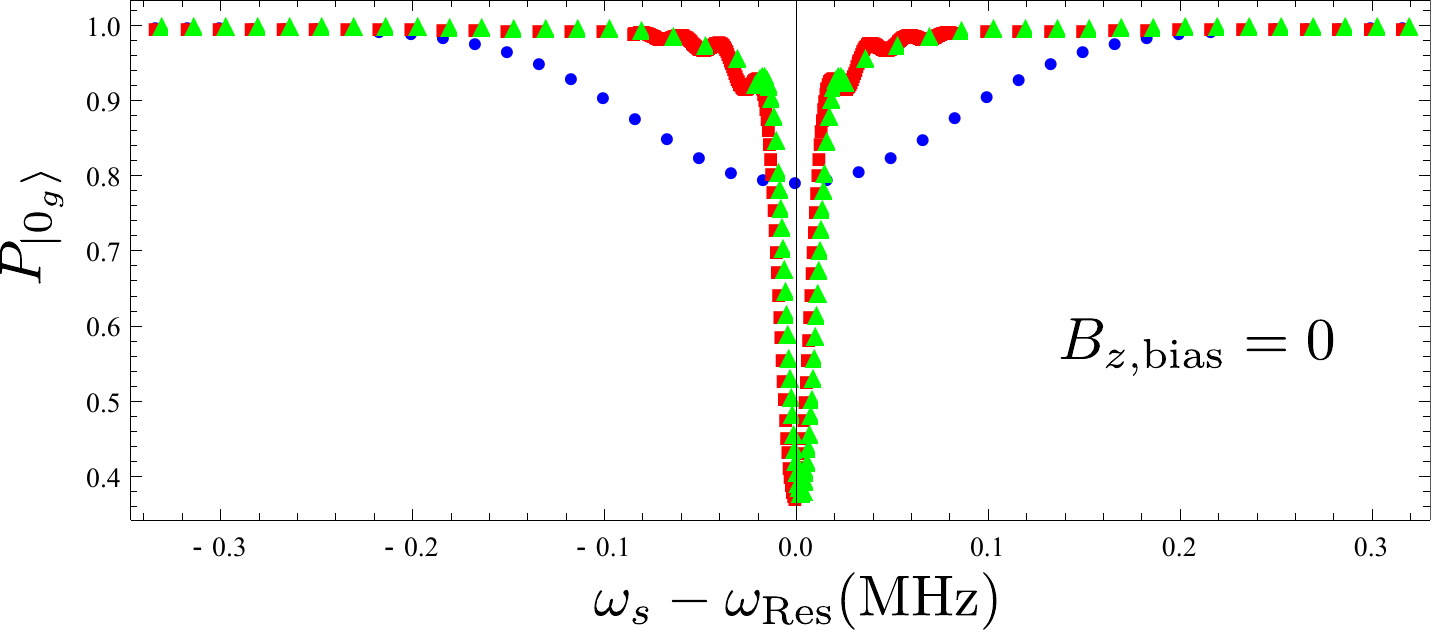}\caption{\label{fig:FigPw0Field}(color online). Magnetic resonance signals for the NV center
in zero field ($B_{z,\text{bias}}=0$) at the moment $t=50~\mu$s, for the cases without optical
control (blue dots) and with optical control ($\Omega=10$~MHz, red
squares). The amplitude of the magnetic source fields $\eta_{0}=0.01$~MHz. The green triangles are the resonance signal for the case where $\Omega_{\pm}$ have independent fluctuations with a relative standard
deviation $\delta_{\Omega} =0.005$. The data were obtained by $10^4$ runs of averaging.}
\end{figure}

When there is no static bias field $(B_{z,\text{bias}}=0)$, $\delta\approx2~\text{GHz}$, the states
$|\pm1_{g}\rangle$ are degenerate, and
\begin{equation}
\tilde{H}_{\text{sig}}=\eta_{\text{sig}}(t)e^{-i\epsilon_{0,-1}t}e^{-i\theta_{\text{sig}}}\frac{1}{\sqrt{2}}\left(|+1_{g}\rangle+|-1_{g}\rangle e^{i2\theta_{\text{sig}}}\right)\langle0_{g}|+\text{H.c.}.\label{eq:HsigB0}
\end{equation}
By choosing the laser phase
$\phi_{L}=2\theta_{\text{sig}}+\pi$, we achieve the largest sensitivity
as the signal Hamiltonian $\tilde{H}_{\text{sig}}=\eta_{\text{sig}}(t)e^{-i\epsilon_{01}t}e^{-i\theta_{\text{sig}}}e^{-i\varphi_{b}}|d_{g}\rangle\langle0_{g}|+\text{H.c.}.$ The states $|d_{g}\rangle$ and $|0_{g}\rangle$ are coherence protected. 
We obtain the magnetic resonance signal by tuning the resonant frequency. 
When the resonant frequency is tuned to the frequency of the signal fields, the state $|0_{g}\rangle$ will transit to $|d_{g}\rangle$ and the change of $P_{|0_{g}\rangle}(t)$ gives the signal.

In Fig.~\ref{fig:FigPw0Field}, we plot the magnetic resonance signal
at the evolution time $t=50$~$\mu$s. The laser phase
is $\phi_{L}=2\theta_{\text{sig}}+\pi$, and the signal fields have
an amplitude $\eta_{0}=0.01$~MHz. Without optical control, the signal
fields with frequency $\omega_{\text{Res}}$ lead to a maximum
resonant population change  $\Delta P_{|0_{g}\rangle}\approx 21\%$
with a linewidth $\Delta\omega_{\text{FWHM}}\approx0.2$~MHz (defined
by the full width at half maximum); while with optical control $\Omega=10$~MHz,
the signal induces a much larger population dip $\Delta P_{|0_{g}\rangle}\approx 63\%$
with a much narrower linewidth $\Delta\omega_{\text{FWHM}}\approx0.02$~MHz,
which is limited by the evolution time $t$ (within the coherence time range). 

To reduce the resonance frequency $\omega_{\text{Res}}$ within the
MHz range, we apply a static bias magnetic field along the axis of
the NV center to narrow the energy gap $\epsilon_{0,-1}$.
When the magnetic field $B_{z,\text{bias}}\approx0.1$~T, $\delta\approx5.7~\text{GHz}$, the ground
states $|0_{g}\rangle$ and $|-1_{g}\rangle$ have an energy gap within the MHz range, and the energy gap between $|0_{g}\rangle$
and $|1_{g}\rangle$ is large ($\gtrsim2.9$ GHz). For large energy
gaps between $|0_{g}\rangle$ and $|1_{g}\rangle$, the transition 
from $|0_{g}\rangle$ to $|1_{g}\rangle$ induced by signal fields is negligible and we have
\begin{equation}
\tilde{H}_{\text{sig}}\approx\eta_{\text{sig}}(t)\frac{1}{2}e^{-i\epsilon_{0,-1}t}e^{i(\theta_{\text{sig}}-\varphi_{b}-\phi_{L})}\left(|b_{g}\rangle-|d_{g}\rangle\right)\langle0_{g}|+\text{H.c.}.\label{eq:HsigBX}
\end{equation}
\begin{figure}
\includegraphics[width=3.4039in]{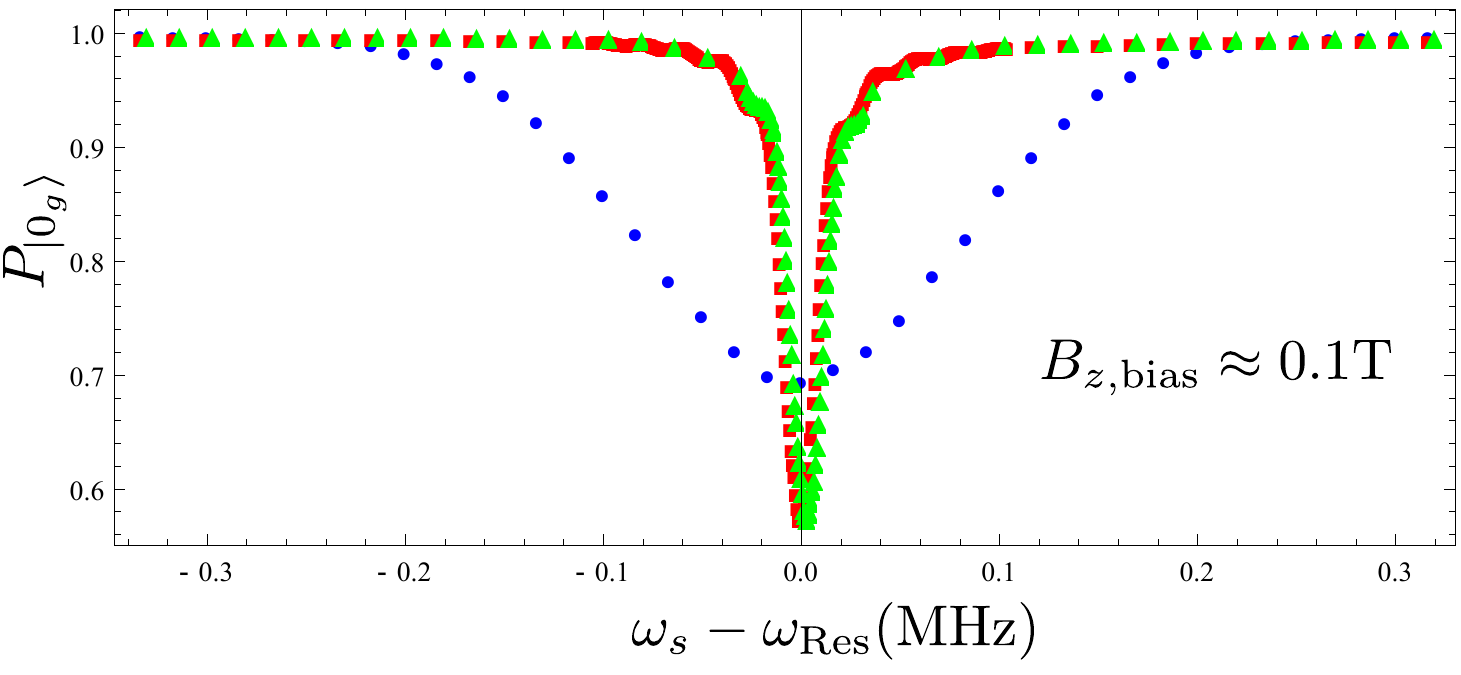}\caption{\label{fig:FigPwXField}(color online). Magnetic resonance signals for the NV center
in a bias field $B_{z,\text{bias}}\approx0.1$~T at the moment $t=50~\mu$s, for the cases
without optical control (blue dots) and with optical control $\Omega=7$~MHz
(red squares). The amplitude of the magnetic source fields $\eta_{0}=0.02$~MHz. The green triangles are the resonance signal for the case where $\Omega_{\pm}$ have independent fluctuations with a relative standard
deviation $\delta_{\Omega} =0.005$. The data were obtained by $10^4$ runs of averaging.}
\end{figure}
Under optical control with a large $\Omega$, we can ensure that the
spectral density of $\beta_{z}(t)$ around $\frac{\Omega}{2}$ and
$\eta_{\text{sig}}(t)$ around $\epsilon_{0,-1}\pm\frac{\Omega}{2}$ is
negligible. The transitions between $|0_{g}\rangle$ and $|b_{g}\rangle$
are suppressed, and we get
\begin{equation}
\tilde{H}_{\text{sig}}\approx-\eta_{\text{sig}}(t)\frac{1}{2}e^{-i\epsilon_{0,-1}t}e^{i(\theta_{\text{sig}}-\varphi_{b}-\phi_{L})}|d_{g}\rangle\langle0_{g}|+\text{H.c.}.
\end{equation}
Therefore, when the field $\eta_{\text{sig}}(t)$ is on resonant with the
transition frequency around $\omega_{\text{Res}}\approx\epsilon_{0,-1}$, the population of $|0_{g}\rangle$
decreases. In this way, the Fourier components of the signal source
$\eta_{\text{sig}}(t)$ can be measured in high resolution. In Fig.~\ref{fig:FigPwXField}, $\eta_{0}=0.02$ MHz, $B_{z,\text{bias}}\approx0.1$~T, and
the evolution time $t=50$~$\mu$s. Without optical
control, the signal fields lead to a maximum population dip $\Delta P_{|0_{g}\rangle}\approx30\%$
with a linewidth $\Delta\omega_{\text{FWHM}}\approx0.2$~MHz; while
with optical control $\Omega=7$~MHz, the population has a larger peak $\Delta P_{|0_{g}\rangle}\approx42\%$
with a much narrower linewidth $\Delta\omega_{\text{FWHM}}\approx0.02$~MHz limited by finite evolution time.

In Figs.~\ref{fig:FigPw0Field} and~\ref{fig:FigPwXField},
we also plot the magnetic resonance signals for optical control
with independent driving fluctuations $\delta\Omega_{\pm}(t)$. It can be seen that an experimentally reachable standard
deviation of relative fluctuations $\delta_{\Omega}=0.005$
only induces tiny changes in the magnetic resonance signal.

\subsection{Measurement of the directions of signal sources}
For the case of $(B_{z,\text{bias}}=0)$, the signal Hamiltonian Eq.~(\ref{eq:HsigB0}) depends on the direction $\theta_{\text{sig}}$ of the signal source. 
Note that the transition from $|0_{g}\rangle$ to $|b_{g}\rangle$
is suppressed when there is optical control. 
If $\phi_{L}=2\theta_{\text{sig}}$ in Eq.~(\ref{eq:HsigB0}), the effect of the signal Hamiltonian
$\tilde{H}_{\text{sig}}=\eta_{\text{sig}}(t)e^{-i\epsilon_{01}t}e^{-i\theta_{\text{sig}}}e^{-i\varphi_{b}}|b_{g}\rangle\langle0_{g}|+\text{H.c.}$
is suppressed by the energy gap between $|0_{g}\rangle$ and $|b_{g}\rangle$
states, and the population change  $\Delta P_{|0_{g}\rangle}$ is small. We have shown that the signal is large when the laser phase
$\phi_{L}=2\theta_{\text{sig}}+\pi$. We use this phase dependence to
determine the direction $\theta_{\text{sig}}$ of the signal field.

\begin{figure}
\includegraphics[width=3.4039in]{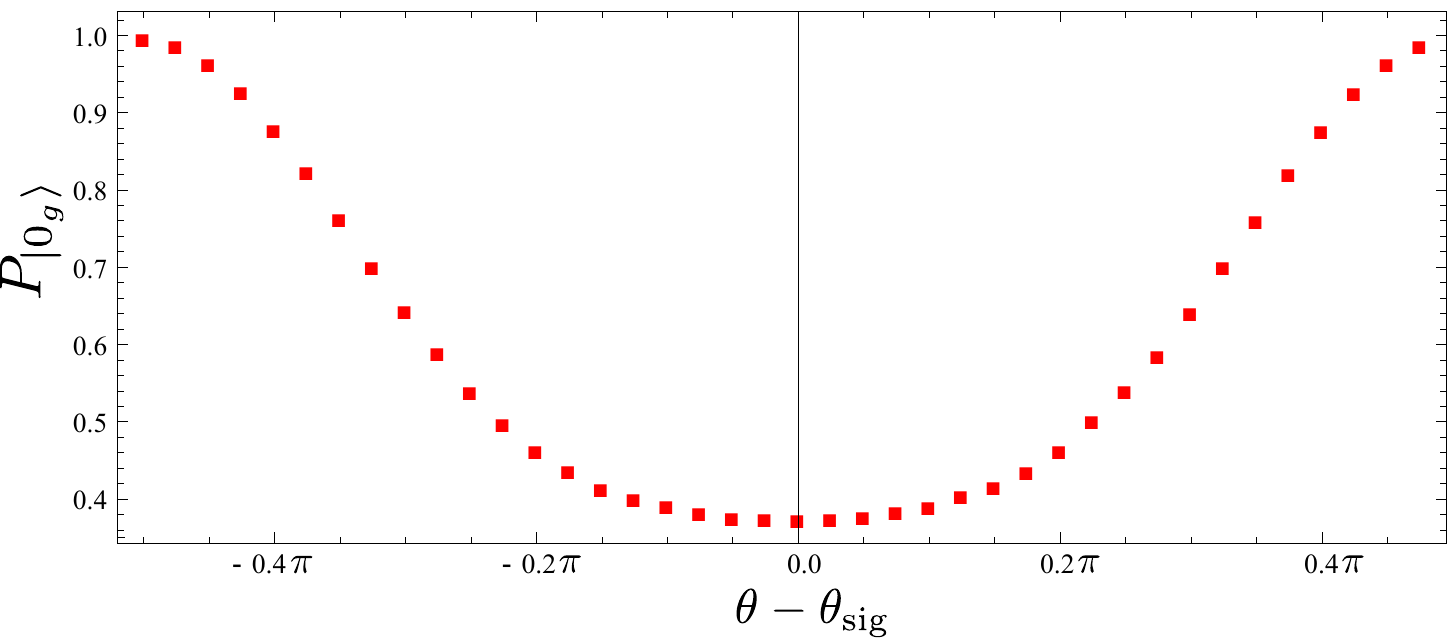}\caption{\label{fig:FigAngle}(color online). The on-resonant ($\omega_{s}=\omega_{\text{Res}}$)
magnetic signal (the change of the population in $|0_{g}\rangle$)
as a function of the angle difference $\theta-\theta_{\text{sig}}$ at the moment $t=50~\mu$s.
The parameters $B_{z,\text{bias}}\approx0$, $\Omega=10$~MHz, and
$\eta_{0}=0.01$~MHz. The data were obtained by $5\times10^{3}$ runs of averaging.}
\end{figure}

In Fig.~\ref{fig:FigAngle},
we applied a resonant signal field at the frequency $\omega_{s}=\omega_{\text{Res}}$ 
and tuned the laser phase $\phi_{L}=2\theta+\pi$ with different angles
$\theta$. The control strength $\Omega=10$~MHz and the amplitude of the signal field $\eta_{0}=0.01$~MHz.
From the lowest point in Fig.~\ref{fig:FigAngle}, we
can infer the direction $\theta_{\text{sig}}$ of the signal source
in the $x-y$ plane. For the case without optical control, $x$ and $y$ directions are equivilent and it is obvious that we cannot infer the direction of the signal source in the $x-y$ plane.
Note that if the laser phase $\phi_{L}$ has a small deviation
in the region $2(\theta_{\text{sig}}-0.1\pi)+\pi\lesssim\phi_{L}\lesssim2(\theta_{\text{sig}}+0.1\pi)+\pi$,
the signal $\Delta P_{|0_{g}\rangle}$ is also large with little
change $(\approx 1\%)$. Therefore, our scheme is robust to fluctuations of laser
phase.

\subsection{Sensitivity enhanced by optical control}

\begin{figure}
\includegraphics[width=3.4039in]{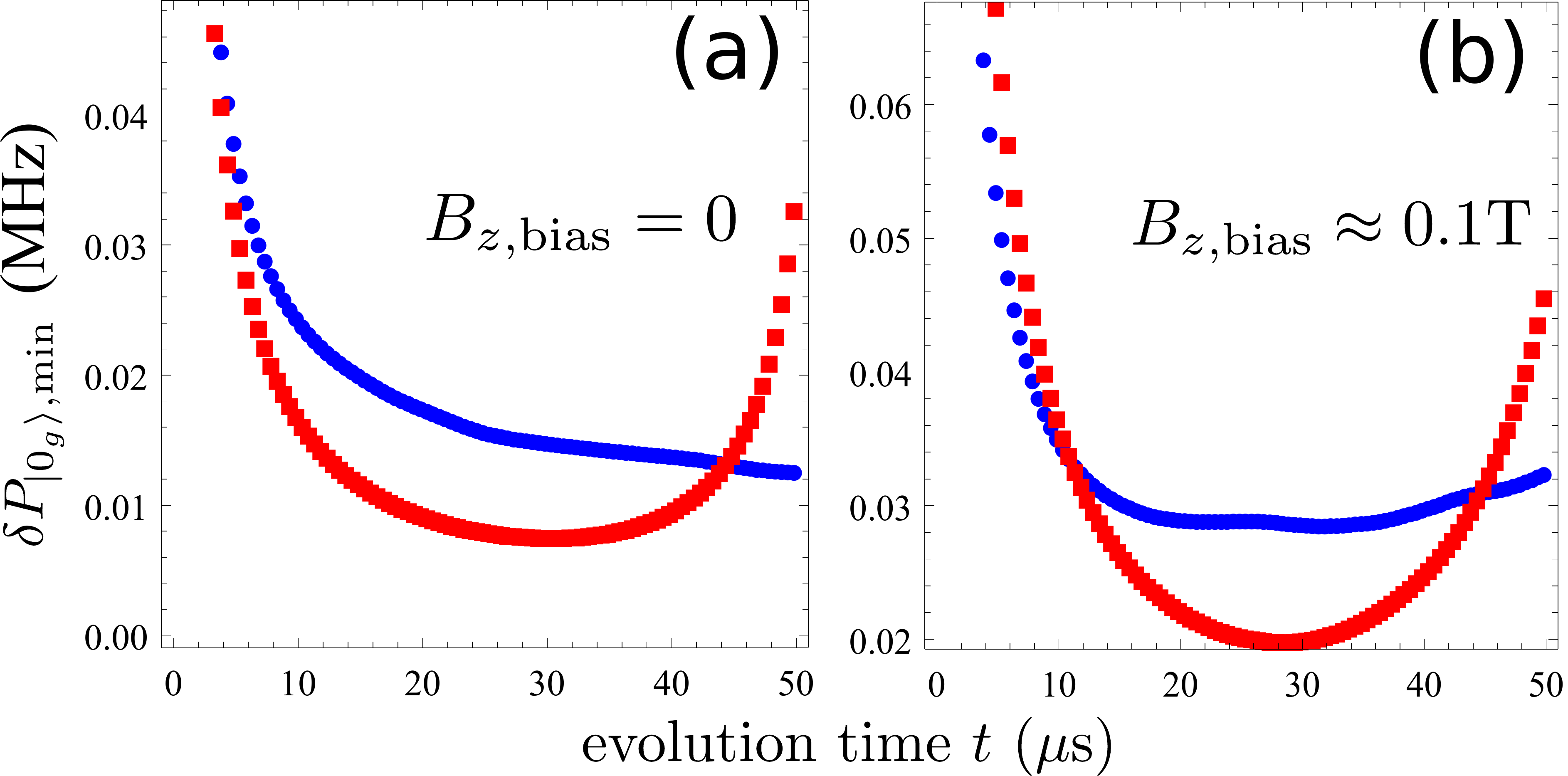}\caption{\label{fig:FigSen}(color online). The one-trial sensitivity of the magnetic resonance signal
at the resonant frequency $\omega_{s}=\omega_{\text{Res}}$ for the
cases of without optical control (blue dots) and with optical control (red squares). (a) $B_{z,\text{bias}}=0$, $\eta_{0}=0.01$~MHz, and $\Omega=10$~MHz;  (b) $B_{z,\text{bias}}\approx0.1$~T, $\eta_{0}=0.02$~MHz, and $\Omega=7$~MHz. The data were obtained by $10^5$ runs of averaging.}
\end{figure}

Note that in Figs.~\ref{fig:FigPw0Field} and \ref{fig:FigPwXField},
the signal peaks with optical control are more pronounced. This implies
that the sensitivity is also improved when the dephasing noise is
suppressed by the optical control. 
In Fig.~\ref{fig:FigSen}, we plot the one-trial sensitivity $\delta P_{|0_{g}\rangle,\text{min}}=\Delta P_{|0_{g}\rangle}/(\frac{\partial P_{|0_{g}\rangle}}{\partial\eta_{0}})$
for the cases of the bias magnetic field $B_{z,\text{bias}}\approx0$
and $B_{z,\text{bias}}\approx0.1$~T, respectively. The parameters
are the same as those for the optical sensing in Figs.~\ref{fig:FigPw0Field}
and \ref{fig:FigPwXField} with $\omega_{s}=\omega_{\text{Res}}$.
Here $\Delta P_{|0_{g}\rangle}=\sqrt{P_{|0_{g}\rangle}(1-P_{|0_{g}\rangle})}$
is the standard deviation of the population $P_{|0_{g}\rangle}$ in one measurement. Averaging the data by repeating
the measurement $N$ times improves the sensitivity by a factor of $\alpha_{s}=1/\sqrt{N}$. If we perform the experiment for a given time $T_{\text{all}}$, we have $N=T_{\text{all}}/(T_{\text{init}}+t)$, where $T_{\text{init}}$ is the time for both initialization and readout and $t$ is the evolution time of the NV spin.
With a large bias field, the transition
from $|0_{g}\rangle$ to $|+1_{g}\rangle$ is suppressed by the energy
gap between $|0_{g}\rangle$ and $|+1_{g}\rangle$ {[}see Eq.~(\ref{eq:HsigBX}){]}.
Therefore the effective signal strength is reduced and the sensitivity
with optical control is reduced in Fig.~\ref{fig:FigPwXField} at
a short sensing time $t\lesssim10$~$\mu$s. At a longer sensing
time, the benefits from decohernce suppression by optical control
manifest and the sensitivity is improved. At zero bias field, the
transition from $|0_{g}\rangle$ to $|+1_{g}\rangle$ is kept under
optical control, and the one-trial sensitivity is improved for a wide range of times
$t\lesssim45$~$\mu$s in the figure. The reduction of the one-trial sensitivity with optical control at $t\gtrsim45$~$\mu$s is caused by the leakage of the population out of the $\Lambda$-type system.

\section{Discussion and conclusion}

We have proposed an all-optical scheme to prolong the quantum coherence
of a negatively charged NV center in diamond. With the quantum coherence
extended and the energy fluctuations in the $^{3}\text{A}_{2}$ ground
sublevels suppressed by the optical driving fields, we have achieved magnetic
resonance with much narrower spectral linewidth and higher detection
sensitivity. Unlike magnetic resonance by pulse sequences or more
generally by dynamical decoupling, in our scheme driving field fluctuations
do not broaden the resonant signal peaks. The sensing frequency by
optical control is determined by the energy gap of the NV $^{3}\text{A}_{2}$
ground sublevels, which can easily reach the GHz range and enables stable
GHz frequency standards in solids. At zero field, the magnetic resonance
spectrum also enables measurement of the direction of signal sources
in the plane perpendicular to the NV symmetry axis. The performance of the all-optical scheme has been confirmed by numerical simulations,  
by selecting the driving amplitudes within a range where the transitions to the off-resonant excited state are small and the $\Lambda$-type optical transition is a good approximation.
Although we apply
the optical driving fields on NV centers in diamond, the method is general
and is applicable to other systems where a $\Lambda$-type optical
transition can be formed.

\begin{acknowledgments}
This work is supported by an Alexander von Humboldt Professorship, the EU Integrating projects SIQS and DIADEMS and the DFG via SPP 1601. A.R. acknowledges the support of ISF grant no. 1281/12.
\end{acknowledgments}



\appendix
\section{\label{app:freqDomain}Decoherence function in the frequency domain}
Here we outline the decoherence suppression by optical control in the frequency domain.
For simplicity, we
consider a simplified model without contributions from spontaneous decay and examine the effect of $\beta_{z}(t)$
up to the second order:
\begin{equation}
L_{0,d_{g}}(t)=1-\frac{1}{2}\int_{0}^{t}dt_{1}\int_{0}^{t_{1}}dt_{2}\overline{\beta_{z}(t_{1})\beta_{z}(t_{2})}M(t_{1},t_{2}),
\end{equation}
where the modulation function
\begin{eqnarray}
M(t_{1},t_{2}) & = & \cos\frac{\Omega t_{1}}{2}\cos\frac{\Omega t_{2}}{2}+\sin\frac{\Omega t_{1}}{2}\sin\frac{\Omega t_{2}}{2},\\
 & = & \cos\left[\frac{\Omega}{2}(t_{1}-t_{2})\right].
\end{eqnarray}
With $M(t_{1},t_{2})=M(t_{2},t_{1})$ and the symmetry $\overline{\beta_{z}(t_{1})\beta_{z}(t_{2})}=\overline{\beta_{z}(t_{2})\beta_{z}(t_{1})}$
for classical noise, we have

\begin{equation}
L_{0,d_{g}}(t)=1-\frac{1}{2}\int_{0}^{t}dt_{1}\int_{0}^{t}dt_{2}\overline{\beta_{z}(t_{1})\beta_{z}(t_{2})}M(t_{1},t_{2}).
\end{equation}
We assume stationary noise; i.e.,  noise with time translation
symmetry, $\overline{\beta_{z}(t_{2})\beta_{z}(t_{1})}=\overline{\beta_{z}(t_{2}-t_{1})\beta_{z}(0)}$.
We write
\begin{equation}
L_{0,d_{g}}(t)=1-\frac{1}{2}\int_{-\infty}^{\infty}\frac{d\omega}{2\pi}S_{\beta}(\omega)\tilde{M}(\omega),
\end{equation}
in terms of the spectral power density
\begin{equation}
S_{\beta}(\omega)=\int_{-\infty}^{\infty}dt\overline{\beta_{z}(t)\beta_{z}(0)}e^{i\omega t},
\end{equation}
and the filter function
\begin{equation}
\tilde{M}(\omega)=\int_{0}^{t}dt_{1}\int_{0}^{t}dt_{2}e^{-i\omega(t_{1}-t_{2})}M(t_{1},t_{2}).
\end{equation}
Without optical control, i.e., $\Omega=0$, the function $\tilde{M}(\omega)=4\sin^{2}\left(\frac{\omega t}{2}\right)/\omega^{2}$
cannot filter out low-frequency fluctuations, which are dominant sources
of decoherence. With large optical driving fields, the filter function $\tilde{M}(\omega)\approx C_{\omega,\Omega}\sin^{2}\left(\frac{\Delta_{\omega,\Omega}t}{2}\right)/\Delta_{\omega,\Omega}^{2}$
with $C_{\omega,\Omega}=4\omega\Omega/(\omega+\frac{\Omega}{2})^{2}$
has a power-law decay with the deviation $\Delta_{\omega,\Omega}=|\omega-\frac{\Omega}{2}|$,
and the low frequency fluctuations are filtered out for large $\Omega$.
In this simplified model, when the spectral power density around the frequency $\frac{\Omega}{2}$
is negligible, $|L_{0,d_{g}}(t)|\approx1$ and the decoherence is
strongly suppressed.

\end{document}